\documentclass[11pt,a4paper]{article}

\usepackage{latexsym,amsmath,amssymb,amsthm,bm,booktabs,multicol,authblk}

\newcommand{\bfx}{\boldsymbol{x}}

\newcommand{\ds}{\displaystyle}

\newtheorem{thm}{Theorem}

\newtheorem{prop}[thm]{Proposition}

\qedhere

\title{Fractional programming formulation for the vertex coloring problem}

\date{\empty}

\author{Tomomi~Matsui\thanks{e-mail: matsui.t.af@m.titech.ac.jp}}
\author{Noriyoshi~Sukegawa\thanks{e-mail: sukegawa.n.aa@m.titech.ac.jp}}
\author{Atsushi~Miyauchi\thanks{e-mail: miyauchi.a.aa@m.titech.ac.jp}}

\affil{{\small \it{Graduate School of Decision Science and Technology, 
Tokyo Institute of Technology, Ookayama 2-12-1, Meguro-ku, Tokyo 152-8552, Japan}}}

\begin{document}
\maketitle
\vspace{-8mm}

\begin{abstract}
We devise a new formulation for the vertex coloring problem. 
Different from other formulations, 
decision variables are associated with the pairs of vertices. 
Consequently, colors will be distinguishable. 
Although the objective function is fractional, 
it can be replaced by a piece-wise linear convex function. 
Numerical experiments show that our formulation has significantly good performance for dense graphs. 
\end{abstract}

\section{Introduction} \label{sec:Intro}

The vertex coloring problem (VCP) is a well-known NP-hard~\cite{GaJo} combinatorial optimization problem 
with a large number of applications including scheduling, register allocation, and timetabling 
(see the survey \cite{MaTo} for the details). 
In this problem, 
we are given a simple and undirected graph $G=(V, E)$. 
The objective is to find an assignment of colors to $V$ such that 
no two adjacent vertices share the same color and 
the number of colors used is minimized. 

\par 

In the standard formulation for VCP, letting $C$ be a set of colors, 
we introduce a decision variable $x_{vc}$ ($\forall v \in V$, $\forall c \in C$) 
which takes 1 if $v$ receives color $c$ and takes 0 otherwise. 
Since every graph can be colored with $n=|V|$ colors, 
it suffices to set $C=\{1,2,\ldots, n\}$. 
Although this formulation is intuitive and simple, 
there exists a strong symmetry in the feasible region 
resulting from the indistinguishability of colors. 
Suppose that we have a solution using $k$ colors. 
Then we see that this model has $\binom{|C|}{k} k!$ equivalent solutions. 
This property will be a great disadvantage when we use ILP solvers. 
For this reason, cuts that remove the symmetry have been studied~\cite{MeZa, MeZa2}. 
On the other hand, recently, alternative formulations for VCP have received a considerable attention. 
For instance, there are studies on a set partitioning formulation~\cite{MeTr}, 
an asymmetric representative formulation~\cite{CaCaCo,CaCoFr}, 
an unconstrained quadratic binary programming formulation~\cite{KoGlAlRe}, and a supernodal formulation~\cite{Bu}. 
For further discussion on these formulations, see Burke et al.~\cite{Bu}. 

\par 

In this study, we focus on the pairs of vertices which can be colored by the same color, 
and associate decision variables with these pairs. 
As a result, we obtain a new formulation for VCP. 
Our model does not suffer the symmetry which is discussed above and has a linear fractional objective function. 
This objective function can be equivalently replaced by a piece-wise linear convex function, 
which gives us a mixed integer linear programming (MILP) model for VCP. 
By this transformation, we can feed our model to commercial MILP solvers such as Gurobi Optimizer. 
To verify the validity of our formulation, we conducted numerical experiments 
on random graphs and several instances form DIMACS Implementation Challenge, 
and confirmed that it has a significantly good performance for dense graphs. 
It should be noted that high edge density does not necessarily make instances easy. 
In fact, there is a dense but hard instance \texttt{DSJC125.9} with only $125$ vertices from DIMACS Implementation Challenge. 
The optimal value of this instance was an open problem until very recently. See Gualandi and Malucelli~\cite{GuMa} for the details.  
We confirmed that our model solves this instance less than only $1$ minute. 


\section{Our formulation}\label{sec:F}

\subsection{Expression as a fractional programming problem}

In our formulation, for each distinct pair of vertices $u$ and $v$, 
we introduce a decision variable $x_{uv}$ 
which takes 1 if $u$ and $v$ share the same color and takes 0 otherwise. 
Clearly, we have $x_{uv} = 0$ for each $\{u,v\} \in E$. 
Here, we use the following inequality constraints 
\begin{equation*}
x_{uv} + x_{vw} - x_{uw} \le 1~~~(\forall u, v, w \in V \mbox{ with } u \not= v, v \not= w, u \not= w) 
\end{equation*}
to obtain an explicit description of the feasible region. 
These inequalities say that if $u$ and $v$ share the same color ($x_{uv}=1$) 
and $v$ and $w$ also share the same color ($x_{vw}=1$), 
then $u$ and $v$ must recieve the same color ($x_{uw}=1$). 
These inequalities are referred to as the triangle inequalities 
studied in Gr\"otschel and Wakabayashi~\cite{GrWa} as facet-defining inequalities for a clique partitioning polytope. 
This relationship is natural because 
if $\bfx$ is a feasible solution for VCP, then a set $E_{\bfx} = \{ \{u,v\} \mid x_{uv}=1 \}$ of edges 
induces a clique partitioning of the complement graph $\overline{G}$ of $G$, and vice versa. 

\par

Next, let us consider how to calculate the number of colors used in $\bfx$, namely the objective value. 
To this aim, we focus on the number of connected components in $(V, E_{\bfx})$. 
It is easy to see that this number equals the desired value. 
For a feasible solution $\bfx$, let us define
\begin{equation*}
f_v(\bfx) = \frac{1}{\ds 1+\sum_{u \in V} x_{uv}} 
\end{equation*}
for each $v \in V$. 
Suppose that a vertex $v$ belongs to a connected component $(V', E')$ with $|V'|=k$ in $(V, E_{\bfx})$. 
Then we have $f_v(\bfx) = 1 / k$ since $V'$ is a clique of $(V, E_{\bfx})$. 
Thus, the sum of $f_v(\bfx)$ over $v \in V'$ equals 1 for each connected component $(V', E')$, 
which implies that the sum of $f_v(\bfx)$ over $v \in V$ gives the number of connected components in $(V, E_{\bfx})$. 
Therefore, we obtain the following proposition. 

\begin{prop}
For a given feasible solution $\bfx$, 
\begin{equation*}
\sum_{v \in V} f_v(\bfx)
\end{equation*}
equals the number of connected components in $(V, E_{\bfx})$, which is the number of colors used in $\bfx$. 
\end{prop}

In sum, our formulation is described as follows: 
\begin{equation*}
\begin{array}{llllllllllll}
\mbox{minimize}		&\ds \sum_{v \in V} f_v(\bfx)						&										\\
\mbox{subject to}	&x_{uv} = 0				&(\forall \{u,v\} \in E),							\\
			&x_{uv} + x_{vw} - x_{uw} \le 1		&(\forall u, v, w \in V \mbox{ with } u \not= v, v \not= w, u \not= w),		\\
			&x_{uv} \in \{0,1 \}			&(\forall u, v \in V \mbox{ with } u \not= v). 					\\
\end{array}
\end{equation*}
It should be noted that there are redundant variables and constraints. 
Suppose that $\{u, v\} \in E$. 
Then, of course, we do not need to prepare the decision variable $x_{uv}$. 
In addition, the transitivity constraints 
$x_{uv} + x_{vw} - x_{uw} \le 1$ is redundant for every $w \in V \setminus \{u, v\}$ 
because it is equivalent to $x_{vw} - x_{uw} \le 1$, 
which is satisfied by any pair of $x_{vw}$ and $x_{uw}$ with $0 \le x_{vw}, x_{uw} \le 1$. 
In our numerical experiments, such redundant variables and constraints are removed. 

\subsection{Expression as a mixed integer linear programming problem}

When implementing our formulation on MILP solvers, 
we substitute a piece-wise linear convex function for 
the fractional objective function. 
For each $v \in V$, 
we introduce a continuous decision variable $f_v$ which equals $f_v(\bfx)$ for a given feasible solution $\bfx$. 
Namely, the objective function will be the sum of $f_v$ over $v \in V$. 
For each $v \in V$ and for each $i \in \{0,1, \ldots, n-1 \}$, 
we add the following linear inequality constraint 
\begin{equation*}\label{f_v}
f_v \ge u_i\left(\sum_{u \in V} x_{uv} \right), 
\end{equation*}
where 
\begin{equation*}
u_i(d) = -\frac{1}{(i+1)(i+2)}(d-i) + \frac{1}{i+1}. 
\end{equation*}

If $d \in \{0,1, \ldots, n-1 \}$, then 
$u_d(d) = 1 / (1+d)$ and $u_d(d)$ is the largest value among $\{ u_k(d) \mid k \in \{0,1, \ldots, n-1 \} \}$. 
Hence, if $\bfx$ is a feasible solution, since $d := \sum_{u \in V} x_{uv}$ is an integer 
and $f_v$ is minimized in the objective function, we have 
\begin{equation*}
f_v = u_d(d) = \frac{1}{1+d} = \frac{1}{\ds 1+\sum_{u \in V} x_{uv}} = f_v(\bfx) 
\end{equation*}
for every $v \in V$. 
This shows the validity of the above constraints. 

\section{Numerical experiments}\label{sec:CompRes}


\begin{table}[tb]
\caption{Results for the randomly generated graphs\label{tbl:smallRand}}
\begin{center}
{\small
\begin{tabular}{rrrrr}														
\toprule
\multicolumn{2}{l}{Instance}			&&	\multicolumn{2}{c}{Our formulation}\\ \cline{1-2} \cline{4-5}
\multicolumn{1}{c}{$n$}&\multicolumn{1}{c}{$p$}&&\multicolumn{1}{l}{Time[s]}&\multicolumn{1}{l}{Gap[\%]}\\
\midrule										
30	&	0.3	&&	3.44		&	---\\	
	&	0.5	&&	4.48		&	---\\
	&	0.7	&&	0.17		&	---\\	
	&	0.9	&&	0.04		&	---\\
\midrule
50	&	0.3	&&	*****	&	16.43	\\	
	&	0.5	&&	**		&	10.06\\
	&	0.7	&&	11.71		&	---\\	
	&	0.9	&&	0.14		&	---\\
\midrule														
70	&	0.3	&&	*****	&	33.00\\	
	&	0.5	&&	*****	&	17.00	\\
	&	0.7	&&	517.87	&	---\\	
	&	0.9	&&	0.40		&	---\\
\bottomrule
\end{tabular}	
}
\end{center}
\end{table}
\begin{table}[tb]
\caption{Results for the randomly generated graphs \label{tbl:middleRand}}
\begin{center}
{\small
\begin{tabular}{rrrrr}														
\toprule
\multicolumn{2}{l}{Instance}			&&	\multicolumn{2}{l}{Our formulation}\\ \cline{1-2} \cline{4-5}
\multicolumn{1}{c}{$n$}	&	\multicolumn{1}{c}{$p$}	&&	\multicolumn{1}{l}{Time[s]}	&	\multicolumn{1}{l}{Gap[\%]}	\\
\midrule										
100	&	0.9	&&	1.50		&	---\\	
150	&	0.9	&&	61.59	&	---\\								
200	&	0.9	&&	*		&	1.59\\	
\bottomrule
\end{tabular}	
}													
\end{center}
\end{table}
\begin{table}[tb]
\caption{Results for middle-sized dense instances from the Second DIMACS Implementation Challenge \label{tbl:middleDMCS}}
\begin{center}
{\small
\begin{tabular}{lrrrrrrrrrrrrrrrrrrrrrr}
\toprule
\multicolumn{3}{l}{Instance}	&&		&&\multicolumn{2}{l}{Best Bounds}		\\	\cline{1-3} \cline{7-8}
\multicolumn{1}{l}{Name}	&\multicolumn{1}{c}{$n$}	&\multicolumn{1}{l}{Density[\%]}
&&\multicolumn{1}{l}{Time[s]}
&&Lower	&Upper	& Opt. \\
\midrule	
\texttt{r125.1c}		&	125	&96.8	&&	0.75		&&	46	&	46 	& 46 \\	
\texttt{DSJC125.9}	&	125	&89.8	&&	57.48		&&	44	&	44 	& 44 \cite{GuMa}\\	
\texttt{DSJC250.9}	&	250	&89.6	&&	7200		&&	71 	&	72 	& 72 \cite{HeCoSe}\\	
\texttt{DSJR500.1c}	&	500	&97.2	&&	7200		&&	79	&	86 	& 85 \cite{GuMa}\\
\bottomrule
\end{tabular}
}
\end{center}
\end{table}


In this section, we report the numerical experiments on our formulation. 
All results were measured on a Linux-based computer with 2.66 GHz quad-core processors and 24 GB RAM. 
We used Gurobi Optimizer 5.6.0 to solve the MILP problems, 
and imposed a time limit of two hours (7200s) on run time per instance. 
In addition, below, ``gap'' means the relative gap, i.e., $(\mbox{upper bound} - \mbox{lower bound}) / \mbox{upper bound}$. 

\par 

The results on random graphs are shown in Table~\ref{tbl:smallRand}. 
We generate five instances for each pair of 
the edge density $p \in \{0.3, 0.5, 0.7, 0.9\} $ and the number of vertices $n \in \{30, 50, 70\}$ 
and exhibit the average computation time for these five instances.  
When calculating the average, 
if there is an instance which cannot be solved to optimality by the limit, 
(since the average does not make sense) 
we count the number of such instances and denote this by the number of ``$*$''s. 
We see that, unfortunately, our formulation fails to solve several instances with $p \le 0.5$ even when $n=50$. 
However, for dense graphs, especially when $p=0.9$, the computation time is very short. 

\par

Motivated by the results in Table~\ref{tbl:smallRand}, 
we further solved three dense and a little bit larger instances. 
These three instances are again randomly generated instances 
where the edge density $p$ is fixed to $0.9$ and the number of vertices $n$ ranges in $\{100, 150, 200\}$. 
Again, for each $n$, we generate five instances. 
We see that, even when $n=150$, optimal solutions are obtained within only about $1$ minute, on average. 
Although, when $n=200$, one instance could not be solved to optimality, the remained relative gap is small. 

\par 

Next, in Table~\ref{tbl:middleDMCS}, 
we show results for several dense and middle-sized instances from the Second DIMACS Implementation Challenge. 
The three instances \texttt{DSJC125.9}, \texttt{DSJC250.9},	and \texttt{DSJR500.1c} are not so large but very hard. 
Indeed, their optimal values were unknown until very recently. See \cite{GuMa} for \texttt{DSJC125.9} and \texttt{DSJR500.1c}, 
and \cite{HeCoSe} for \texttt{DSJC250.9}. 
Although we could not solve the last two instances within two hours, 
the first two instances are solved to optimality. 
Moreover, its computation time is short. 


\section{Conclusion}

In this study, we devise a new formulation for VCP. 
Different from other existing formulations, 
we associate decision variables with the pairs of vertices, 
which makes the colors distinguishable. 
In our formulation, 
the number of colors used is expressed as a linear fractional function. 
To implement the formulation on MILP solvers, 
we replace the fractional objective function by a piece-wise linear convex function. 
The noteworthy property of our formulation is 
its considerably high performance for dense graphs. 


\bibliographystyle{abbrv}
\bibliography{references.bib}

\section*{Adding simple cuts}

In this appendix, we introduce simple valid inequalities to strengthen the formulation. 
Let $|I_v|$ be the size of a maximum independent set $I_v$ which includes $v \in V$ in $G$. 
Then, it is easy to see that the number of vertices which can receive the same color as that of $v$ is 
bounded above by $(|I_v| - 1)$ in any feasible solution. In other words, for each $v \in V$, 
the following inequality 
\begin{equation*}
\sum_{u \in V} x_{uv} \le |I_v| - 1
\end{equation*}
can be added to our formulation. 
Table~\ref{tbl:middleRandCuts} and Table~\ref{tbl:middleDMCSCuts} show 
the results of our formulation with these simple cuts for the instances shown in 
Table~\ref{tbl:middleRand} and Table~\ref{tbl:middleDMCS}, respectively. 
The computation time includes the preprocessing time, 
namely the computation time for calculating $|I_v|$ for each $v \in V$. 
This preprocessing time is exhibited in the brackets. 
Although we also conducted numerical experiments on the instances shown in Table~\ref{tbl:smallRand}, 
the improvement on solving time is not significant, hence, we omit the corresponding results. 
From Table~\ref{tbl:middleRandCuts} and Table~\ref{tbl:middleDMCSCuts}, 
we see that simple cuts provide an advantage on solving time for all instances other than \texttt{DSJC250.9}.  
For \texttt{DSJC250.9}, the upper bound becomes slightly worse. 

\begin{table}[tb]
\caption{Results for the randomly generated graphs with simple cuts \label{tbl:middleRandCuts}}
\begin{center}
{\small
\begin{tabular}{rrrrrrrrrrrrrrrrrrrrrrrr}														
\toprule													
\multicolumn{2}{l}{Instance}			&&	\multicolumn{2}{l}{With simple cuts}	\\	\cline{1-2} \cline{4-5}
\multicolumn{1}{c}{$n$}	&	\multicolumn{1}{c}{$p$}	&&	\multicolumn{1}{l}{Time[s]}	&	\multicolumn{1}{l}{Gap[\%]}	\\
\midrule														
100	&	0.9	&&	1.26	(0.22)		&	---\\	
150	&	0.9	&&	59.97 (0.67)		&	---\\	
200	&	0.9	&&	1587.26 (1.54)		&	---\\
\bottomrule					
\end{tabular}	
}													
\end{center}
\end{table}

\begin{table}[tb]
\caption{Results for middle-sized dense instances from the Second DIMACS Implementation Challenge with simple cuts \label{tbl:middleDMCSCuts}}
\begin{center}
{\small
\begin{tabular}{lrrrrrrrrrrrrrrrrrrrrrr}
\toprule
\multicolumn{3}{l}{Instance}	&&		&&\multicolumn{2}{l}{Best Bounds}		\\	\cline{1-3} \cline{7-8}
\multicolumn{1}{l}{Name}	&\multicolumn{1}{c}{$n$}	&\multicolumn{1}{l}{Density[\%]}
&&\multicolumn{1}{l}{Time[s]}
&&Lower	&Upper	\\
\midrule
\texttt{r125.1c}		&	125	&96.8	&&	0.27 (0.09)			&&	46	&	46 	\\	
\texttt{DSJC125.9}	&	125	&89.8	&&	33.63	 (0.40)		&&	44	&	44 	\\	
\texttt{DSJC250.9}	&	250	&89.6	&&	7200 (3.50)		&&	71 	&	73 	\\	
\texttt{DSJR500.1c}	&	500	&97.2	&&	7200 (1.86)		&&	79	&	85 	\\
\bottomrule
\end{tabular}
}
\end{center}
\end{table}

\end{document}